\documentclass{amsart}
\usepackage{graphicx}
\usepackage{amsmath}
\usepackage{amsfonts}
\usepackage{amssymb}

\begin{document}
\title{Geometry of Entangled States , Bloch Spheres and Hopf Fibrations }
\author{Remy Mosseri$^{1}$ and Rossen Dandoloff$^{2}$}
\address{$^{1}$ Groupe de Physique des Solides, CNRS UMR 7588,\\
Universit\'{e}s Pierre et Marie Curie Paris 6 et Denis Diderot Paris 7,\\
2 place Jussieu, 75251 Paris Cedex 05 France\\
$^{2}$Laboratoire de physique theorique et modelisation, (CNRS-ESA 8089 ),
Universite de Cergy-Pontoise\\
Site de Neuville 3, 5 Mail Gay-Lussac, F-95031 Cergy-Pontoise Cedex}
\email{mosseri@gps.jussieu.fr; rossen.dandoloff@ptm.u-cergy.fr }
\maketitle
\begin{abstract}We discuss a generalization to 2 qubits of the standard Bloch sphere
representation for a single qubit, in the framework of Hopf fibrations of high
dimensional spheres by lower dimensional spheres. The single qubit Hilbert
space is the 3-dimensional sphere $S^{3}$. The $S^{2}$ base space of a
suitably oriented $S^{3}$ Hopf fibration is nothing but the Bloch sphere,
while the circular fibres represent the qubit overall phase degree of freedom.
For the two qubits case, the Hilbert space is a 7-dimensional sphere $S^{7}$,
which also allows for a Hopf fibration, with $S^{3}$ fibres and a $S^{4}$
base. A main striking result is that suitably oriented $S^{7}$ Hopf fibrations
are entanglement sensitive. The relation with the standard Schmidt
decomposition is also discussed
\end{abstract}

\section{ Introduction}

The interest in two-level systems, or coupled two-level systems, is as old as
quantum mechanics itself, with the analysis of the electron spin sector in the
helium atom. The ubiquitous two-level systems have gained a renewed interest
in the past ten years, owing to the fascinating perspectives in quantum
manipulation of information \ and quantum computation\cite{q_info_gen}. These
two-level quantum systems are now called qubits, grouped and coupled into
q-registers, and manipulated by sophisticated means. It is of interest to
describe their quantum evolution in a suitable representation space, in order
to get some insight into the subtleties of this complicated problem. A well
known tool in quantum optics is the Bloch sphere representation, where the
simple qubit state is faithfully represented, up to its overall phase, by a
point on a standard sphere $S^{2}$, whose coordinates are expectation values
of physically interesting operators for the given quantum state. Our aim here
is to build an adequate representation space for bi-partite systems. We are
guided by the relation of the standard Bloch sphere to a geometric object
called the Hopf fibration of the $S^{3\text{ }}$hypersphere (identified to the
spin
${\frac12}$%
Hilbert space). Since the two qubits Hilbert space is the 7-dimensional
sphere\ $S^{7}$, which also allows for a Hopf fibration, it is tempting to
mimic the Bloch sphere representation in that case. An \textit{a priori}
unexpected result is that the $S^{7}$ Hopf fibration is entanglement sensitive
and therefore provides a kind of ''stratification'' \ for the 2 qubits states
with respect to their entanglement content.

This paper is organised as follows. We first briefly recall known facts about
the Bloch sphere representation and its relation to the $S^{3}$ Hopf
fibration. We then proceed to the 2-qubits states and recall what entanglement
and Schmidt decomposition consist in. The $S^{7}$ Hopf fibration is then
introduced and related to the 2-qubits Hilbert space. Although several papers
have appeared (especially in the recent period) which aimed at a geometrical
analysis of qubits Hilbert space \cite{2qbits_geometry}, we are not aware of
an alternative use of the $S^{7}$ Hopf fibration in that context. As far as
computation is concerned, going from the $S^{3}$ to the $S^{7}$ fibration
merely amounts to replacing complex numbers by quaternions. This is why we
give in appendix a brief introduction to quaternion numbers. Note that using
quaternions is not strictly necessary here, but they provide an elegant way to
put the calculations into a compact form, and have (by nature) an easy
geometrical interpretation

\bigskip

\section{ Single qubit, Bloch sphere and the $S^{3}$ Hopf Fibration}

\subsection{The Bloch sphere representation}

A (single) qubit state reads
\begin{equation}
\left|  \Psi\right\rangle =\alpha\left|  0\right\rangle +\beta\left|
1\right\rangle ,\qquad\alpha,\beta\in\mathbb{C}\mathbf{,\qquad}\left|
\alpha\right|  ^{2}+\left|  \beta\right|  ^{2}=1 \label{one-qbit state}%
\end{equation}

In the spin
${\frac12}$%
context, the orthonormal basis $\left\{  \left|  0\right\rangle ,\left|
1\right\rangle \right\}  $ is composed of the two eigenvectors of the (say)
$\sigma_{z}$ (Pauli spin) operator. A conveniant way to represent $\left|
\Psi\right\rangle $ (up to a global phase) is then provided by the Bloch
sphere. The set of states $\exp i\varphi\left|  \Psi\right\rangle (\varphi
\in\left[  0,2\pi\right[  )$ is mapped onto a point on $S^{2}$ (the usual
sphere in \textbf{R}$^{3})$ with coordinates ($X,Y,Z)$%
\begin{align}
X &  =\left\langle \sigma_{x}\right\rangle _{\Psi}=2\operatorname{Re}%
(\overline{\alpha}\beta)\\
Y &  =\left\langle \sigma_{y}\right\rangle _{\Psi}=2\operatorname{Im}%
(\overline{\alpha}\beta)\nonumber\\
Z &  =\left\langle \sigma_{z}\right\rangle _{\Psi}=\left|  \alpha\right|
^{2}-\left|  \beta\right|  ^{2}\nonumber
\end{align}

where $\overline{\alpha}$ is the complex conjugate of $\alpha$. Recall also
the relation between Bloch sphere coordinates and the pure state density
matrix $\rho_{\left|  \Psi\right\rangle }$:
\begin{equation}
\rho_{\left|  \Psi\right\rangle }=\rho_{\exp i\varphi\left|  \Psi\right\rangle
}=\left|  \Psi\right\rangle \left\langle \Psi\right|  =\frac{1}{2}\left(
\begin{array}
[c]{cc}%
1+Z & X-iY\\
X+iY & 1-Z
\end{array}
\right)
\end{equation}

For mixed states, the density matrices are in one-to-one correspondance with
points in the Bloch ''ball'', the interior of the pure state Bloch sphere.

The single qubit Hilbert space is the unit sphere $S^{3}$ embedded in $R^{4}$.
Indeed, writing $\alpha=x_{1}+ix_{2}$, and $\beta=x_{3}+ix_{4}$, the state
normalisation condition translates into $\sum x_{l}^{2}=1$ defining the
$S^{3}$ sphere. In this space, the set of states $\exp i\varphi\left|
\Psi\right\rangle $ is a circle parametrized by $\varphi$. The projective
Hilbert space is such that all states differing by a global phase are
identified, and therefore corresponds to the above Bloch sphere. It can also
be identified to the Hopf fibration basis as follows now.

\subsection{ The $S^{3}$ Hopf fibration}

Briefly said, a space is fibred if it has a sub-space (the fibre) which can be
shifted by a displacement, so that any point of the space belongs to one, and
only one, fibre. For example, the euclidean space $R^{3}$ can be seen as a
fibre bundle of parallel straight lines, but also of parallel planes (the
fibre need not to be one-dimensional). More precisely, a fibred space $E$ is
defined by a (many-to-one) map from $E$ to the so-called ''base space'' $B$,
all points of a given fibre $F$ being mapped onto a single base point. In the
preceeding $R^{3}$ example, the base is either a plane cutting the whole set
of parallel lines, or, in the second case, a line cutting the set of parallel
planes. We are facing \ here ''trivial fibrations'', in the sense that the
base $B$ is embedded in the fibred space $E$ , the latter being faithfully
described as the direct product of the base and the fibre: $R^{3}=R^{2}\times
R$ or $R^{3}=R\times R^{2}$.

The simplest, and most famous, example of a non trivial fibration is the Hopf
fibration of $S^{3}$ by great circles $S^{1}$ and base space $S^{2}$. For the
qubit Hilbert space purpose, the fibre represents the global phase degree of
freedom, and the base $S^{2}$ is identified as the Bloch sphere. The standard
notation for a fibred space is that of a map $E\overset{F}{\rightarrow}B$,
which here reads $S^{3}\overset{S^{1}}{\rightarrow}S^{2}$. Its non trivial
character implies that $S^{3}\neq S^{2}\times S^{1}$. This non trivial
character translates here into the known failure in ascribing consistantly a
definite phase to each representing point on the Bloch sphere.

To describe this fibration in an analytical form, we go back to the definition
of $S^{3}$ as pairs of complex numbers $\left(  \alpha,\beta\right)  $ which
satisfy $\left|  \alpha\right|  ^{2}+\left|  \beta\right|  ^{2}=1$. The Hopf
map is defined as the composition of a map $h_{1}$ from $S^{3}$ to $R^{2}$
$(+\infty)$, followed by an inverse stereographic map from $R^{2}$ to $S^{2}:$%

\begin{align}
h_{1}  &  :%
\begin{array}
[c]{ccc}%
S^{3} & \longrightarrow &  R^{2}+\left\{  \infty\right\} \\
\left(  \alpha,\beta\right)  & \longrightarrow &  C=\overline{\alpha\beta
^{-1}}%
\end{array}
\qquad\alpha,\beta\in\mathbb{C}\nonumber\\
h_{2}  &  :%
\begin{array}
[c]{ccc}%
R^{2}+\left\{  \infty\right\}  & \longrightarrow &  S^{2}\\
C & \longrightarrow &  M(X,Y,Z)
\end{array}
\qquad X^{2}+Y^{2}+Z^{2}=1
\end{align}

The first map $h_{1}$ clearly shows that the full $S^{3}$ great circle,
parametrized by ($\alpha\exp i\varphi,\beta\exp i\varphi$) is mapped on the
same single point with complex coordinate $C$. Note that the complex
conjugation, in the above definition for the Hopf map h$_{1}$, is not
necessary to represent a great circle fibration. It is used here on purpose to
get an exact one-to-one relation with the above Bloch sphere coordinates, and
to prepare the generalization to higher dimension discussed in the next
paragraph.\ It is indeed a simple exercise to show that, with $R^{2}$ cutting
the unit radius $S^{2}$ along the equator, and the north pole as the
stereographic projection pole, the $S^{2}$ Hopf fibration base coordinates
coincide with the above $S^{2}$ Bloch sphere coordinates.

Although there is nothing really new in this correspondance \cite{urtbanke},
it is probably poorly known in both communities (quantum optics and geometry).
It is striking that the simplest non trivial object of quantum physics, the
two-level system, bears such an intimate relation with the simplest non
trivial fibred space.

It is tempting to try to visualize the full ($S^{3}$) Hilbert space with its
fibre structure. This can be achieved by doing a (direct) stereographic map
from $S^{3}$ to $R^{3}$ (nice pictures can be found in \cite{urtbanke}%
,\cite{sadoc_mosseri_livre}). Each $S^{3\text{ }}$circular fibre is mapped
onto a circle in $R^{3}$, with an exceptional straight line, image of the
unique $S^{3}$ great circle passing through the projection pole. The great
circle arrangement is intricate, since each circular fibre threads all others
in $S^{3}$.

\section{Two qubits, entanglement and the $S^{7}$ Hopf fibration}

\subsection{The two-qubit Hilbert space}

We now proceed one step further, and investigate pure states for two qubits.
The Hilbert space $\mathcal{E}$ for the compound system is the tensor product
of the individual Hilbert spaces $\mathcal{E}_{1}\otimes$ $\mathcal{E}_{2}$,
with a direct product basis $\left\{  \left|  00\right\rangle ,\left|
01\right\rangle ,\left|  10\right\rangle ,\left|  11\right\rangle \right\}  $.
A two qubit pure state reads
\begin{align}
\left|  \Psi\right\rangle  &  =\alpha\left|  00\right\rangle +\beta\left|
01\right\rangle +\gamma\left|  10\right\rangle +\delta\left|  11\right\rangle
\qquad\label{two-qbit state}\\
\text{with \ }\alpha,\beta,\gamma,\delta &  \in\mathbb{C},\;\text{and }\left|
\alpha\right|  ^{2}+\left|  \beta\right|  ^{2}+\left|  \gamma\right|
^{2}+\left|  \delta\right|  ^{2}=1\nonumber
\end{align}

$\left|  \Psi\right\rangle $ is said ''separable'' if, at the price of
possible basis changes in $\mathcal{E}_{1}$ and $\mathcal{E}_{2}$ separately,
it can be written as a single product. As is well known, the separability
condition reads: $\alpha\delta=\beta\gamma$. A generic state is not separable,
and is said to be ''entangled''. The Schmidt decomposition \cite{peres}
implies that a suitable basis exists such that $\left|  \Psi\right\rangle $
can be written as a sum of only two (bi-orthonormal) terms with real positive
coefficients. Note that the apparently obvious decomposition
\begin{align}
\left|  \Psi\right\rangle  &  =\cos\Omega\left(  \left|  0\right\rangle
_{1}\otimes\left|  u_{2}\right\rangle _{2}\right)  +\sin\Omega\left(  \left|
1\right\rangle _{1}\otimes(\left|  v_{2}\right\rangle _{2}\right)  \nonumber\\
\text{with }\left|  u_{2}\right\rangle _{2} &  =\frac{1}{\cos\Omega}%
(\alpha\left|  0\right\rangle _{2}+\beta\left|  1\right\rangle _{2})\text{ ,
}\left|  v_{2}\right\rangle _{2}=\frac{1}{\sin\Omega}(\gamma\left|
0\right\rangle _{2}+\delta\left|  1\right\rangle _{2}),\\
\cos\Omega &  =\left(  \left|  \alpha\right|  ^{2}+\left|  \beta\right|
^{2}\right)  ^{%
\frac12
}\text{ and }\sin\Omega=\left(  \left|  \gamma\right|  ^{2}+\left|
\delta\right|  ^{2}\right)  ^{%
\frac12
},\quad\Omega\in\left[  0,\pi/2\right]  \nonumber
\end{align}

is generically not a Schmidt decomposition since $\left\langle u_{2}\mid
v_{2}\right\rangle $ need not vanish, a point on which we return below.

The $\left|  \Psi\right\rangle $ normalization condition $\left|
\alpha\right|  ^{2}+\left|  \beta\right|  ^{2}+\left|  \gamma\right|
^{2}+\left|  \delta\right|  ^{2}=1$ identifies $\mathcal{E}$ to the
7-dimensional sphere $S^{7}$, embedded in $R^{8}$. It is therefore tempting to
see whether the known $S^{7}$ Hopf fibration (with fibres $S^{3}$ and base
$S^{4})$ can play any role in the Hilbert space description.

\subsection{The $S^{7}$ Hopf fibration}

The simplest way to introduce this fibration is to proceed along the same line
as for the $S^{3}$ case, but using quaternions instead of complex numbers (see
appendix). We write
\begin{equation}
q_{1}=\alpha+\beta\mathbf{j},\;q_{2}=\gamma+\delta\mathbf{j},\qquad
q_{1,}q_{2}\in\mathbb{Q}\mathbf{,}%
\end{equation}

and a point (representing the state $\left|  \Psi\right\rangle $) on the unit
radius $S^{7}$ as a pair of quaternions $\left(  q_{1,}q_{2}\right)  $
satisfying $\left|  q_{1}\right|  ^{2}+\left|  q_{2}\right|  ^{2}=1$. The Hopf
map from $S^{7}$ to the base $S^{4}$ is the composition of a map $h_{1}$ from
$S^{7}$ to $R^{4}$ $(+\infty)$, followed by an inverse stereographic map from
$R^{4}$ to $S^{4}$\cite{sadoc_mosseri_livre}.
\begin{align}
h_{1}  &  :%
\begin{array}
[c]{ccc}%
S^{7} & \longrightarrow &  R^{4}+\left\{  \infty\right\} \\
\left(  q_{1},q_{2}\right)  \;\; & \longrightarrow &  Q=\overline{q_{1}%
q_{2}^{-1}}%
\end{array}
\qquad q_{1,}q_{2}\in\mathbb{Q}\nonumber\\
h_{2}  &  :%
\begin{array}
[c]{ccc}%
R^{4}+\left\{  \infty\right\}  & \longrightarrow &  S^{4}\\
Q & \longrightarrow &  M(x_{l})
\end{array}
\qquad\sum\limits_{l=0}^{l=4}x_{l}^{2}=1
\end{align}

This Hopf fibration therefore reads $S^{7}\overset{S^{3}}{\rightarrow}S^{4}$.
The base space $S^{4}$ is not embedded \ in $S^{7}$ : the fibration is again
not trivial. The fibre is a unit $S^{3}$ sphere as can easily be seen by
noting that the $S^{7}$ points $\left(  q_{1,}q_{2}\right)  $ and $\left(
\,q_{1}q,\,q_{2}q\right)  $, with $q$ a unit quaternion (geometrically a
$S^{3}$ sphere) are mapped onto the same $Q$ value. A main difference with the
previous $S^{3}$ Hopf fibration comes from the non commutative nature of the
quaternion product. The Hopf map $h_{1}^{\prime}$, defined by $Q=\overline
{q_{2}^{-1}q_{1}}$ is distinct from $h_{1}$, while still corresponding to a
fibration $S^{7}\overset{S^{3}}{\rightarrow}S^{4}$. One must stress that, due
to the tensor product nature of the two-qubits Hilbert space, the latter
inherits a form of ''anisotropy'' (in $S^{7}$) which translates into specific
choices for the definition of the pair $\left(  q_{1,}q_{2}\right)  $, as will
be discussed below.

Let us now look closer to the base space $S^{4}$ , by working out explicitely
the Hopf map. The $h_{1}$ map leads to%
\begin{align}
Q  &  =\overline{q_{1}q_{2}^{-1}}=\frac{1}{\sin^{2}\Omega}\left[
\overline{\left(  \alpha+\beta\mathbf{j}\right)  \left(  \overline{\gamma
}-\delta\mathbf{j}\right)  }\right]  =\frac{1}{\sin^{2}\Omega}\left(
C_{1}+C_{2}\mathbf{j}\right) \\
\text{with }C_{1}  &  =\left(  \overline{\alpha}\gamma+\overline{\beta}%
\delta\right)  ,\text{ }C_{2}=\left(  \alpha\delta-\beta\gamma\right)  \text{
and }C_{1},C_{2}\in\mathbb{C}\nonumber
\end{align}

Now comes a first striking result: the Hopf map is entanglement sensitive!
Indeed, separable states satisfy $\alpha\delta=\beta\gamma$ and therefore map
onto the subset of pure complex numbers in the quaternion field (both being
completed by $\infty$ when $\sin\Omega=0$). Geometrically, this means that
non-entangled states map from $S^{7}$ onto a 2-dimensional planar subspace of
the target space $R^{4}$. Note however that the latter property heavily
depends on the chosen analytical form of the $h_{1}$ map (in other words, on
the ''orientation'' of the fibres with respect to $S^{7}$). A second
interesting point concerns states whose Hopf map is confined in the
2-dimensional subspace orthogonal to the previous one (such that $C_{1}=0$).
This plane gathers the images of states with the above trivial Schmidt
decomposition: indeed, $C_{1}=0$ implies $\left\langle u_{2}\mid
v_{2}\right\rangle =0.$

The second map $h_{2\text{ }}$sends states onto points on $S^{4}$, with
coordinates $x_{l}$, with $l$ running from to $0$ to $4$. With the inverse
stereographic pole located on the $S^{4}$ ''north'' pole ($x_{0}=+1$), and the
target space $R^{4\text{ }}$ cutting $S^{4}$ along the equator, we get the
following coordinate expressions%
\begin{align}
x_{0}  &  =\cos2\Omega=\left|  q_{1}\right|  ^{2}-\left|  q_{2}\right|  ^{2}\\
x_{1}  &  =\sin2\Omega\;S(Q^{\prime})=2\operatorname{Re}\left(  \overline
{\alpha}\gamma+\overline{\beta}\delta\right) \nonumber\\
x_{2}  &  =\sin2\Omega\;V_{\mathbf{i}}(Q^{\prime})=2\operatorname{Im}\left(
\overline{\alpha}\gamma+\overline{\beta}\delta\right) \nonumber\\
x_{3}  &  =\sin2\Omega\;V_{\mathbf{j}}(Q^{\prime})=2\operatorname{Re}\left(
\alpha\delta-\beta\gamma\right) \nonumber\\
x_{4}  &  =\sin2\Omega\;V_{\mathbf{k}}(Q^{\prime})=2\operatorname{Im}\left(
\alpha\delta-\beta\gamma\right) \nonumber
\end{align}

with $\sin2\Omega=2\left|  q_{1}\right|  \left|  q_{2}\right|  $. $Q^{\prime}$
is the normalized image of the $h_{1}$ map $\ (Q^{\prime}=Q/\left|  Q\right|
)$, $S(Q^{\prime})$ and $V_{\mathbf{i},\mathbf{j},\mathbf{k}}(Q^{\prime})$
being respectively the scalar and vectorial parts of $Q^{\prime}$ (see
appendix). We recalled above that the $X,Y$ and $Z$ Bloch sphere coordinates
are expectation values of \ the spin operators $\sigma$ in the one-qubit state
$\left|  \Psi\right\rangle \left(  \text{eq.}\ref{one-qbit state}\right)  $.
It happens that the $x_{l}$ coordinates are also expectation values of simple
operators in the two-qubits state $\left(  \text{eq.}\ref{two-qbit
state}\right)  $

An obvious one is $x_{0}$ which corresponds to $\left\langle \sigma_{z}\otimes
Id\right\rangle _{\Psi}$. Indeed,%
\begin{align}
\left\langle \sigma_{z}\otimes Id\right\rangle _{\Psi}  &  =\left(
\overline{\alpha},\overline{\beta},\overline{\gamma},\overline{\delta}\right)
\left(
\begin{array}
[c]{cccc}%
1 & 0 & 0 & 0\\
0 & 1 & 0 & 0\\
0 & 0 & -1 & 0\\
0 & 0 & 0 & -1
\end{array}
\right)  \left(
\begin{array}
[c]{c}%
\alpha\\
\beta\\
\gamma\\
\delta
\end{array}
\right) \\
&  =\left|  q_{1}\right|  ^{2}-\left|  q_{2}\right|  ^{2}=\cos2\Omega
=x_{0}\nonumber
\end{align}

The two next coordinates are also easily recovered as
\begin{align}
x_{1}  &  =\left\langle \sigma_{x}\otimes Id\right\rangle _{\Psi}=\left(
\overline{\alpha},\overline{\beta},\overline{\gamma},\overline{\delta}\right)
\left(
\begin{array}
[c]{cccc}%
0 & 0 & 1 & 0\\
0 & 0 & 0 & 1\\
1 & 0 & 0 & 0\\
0 & 1 & 0 & 0
\end{array}
\right)  \left(
\begin{array}
[c]{c}%
\alpha\\
\beta\\
\gamma\\
\delta
\end{array}
\right)  =2\operatorname{Re}\left(  \overline{\alpha}\gamma+\overline{\beta
}\delta\right) \\
x_{2}  &  =\left\langle \sigma_{y}\otimes Id\right\rangle _{\Psi}=\left(
\overline{\alpha},\overline{\beta},\overline{\gamma},\overline{\delta}\right)
\left(
\begin{array}
[c]{cccc}%
0 & 0 & -i & 0\\
0 & 0 & 0 & -i\\
i & 0 & 0 & 0\\
0 & i & 0 & 0
\end{array}
\right)  \left(
\begin{array}
[c]{c}%
\alpha\\
\beta\\
\gamma\\
\delta
\end{array}
\right)  =2\operatorname{Im}\left(  \overline{\alpha}\gamma+\overline{\beta
}\delta\right)
\end{align}

The remaining two coordinates, $x_{3}$ and $x_{4}$, are also recovered as
expectation values of an operator acting on $\mathcal{E}$, but in a more
subtle way. Define $\mathbf{J}$ as the (antilinear) ''conjugator'', an
operator which takes the complex conjugate of all complex numbers involved in
an expression (here acting on the left in the scalar product below). Form then
the antilinear operator $\mathbf{E}$ (for ''entanglor''): $\mathbf{E}%
=-\mathbf{J}\left(  \sigma_{y}\otimes\sigma_{y}\right)  $. One finds%
\begin{align}
x_{3} &  =\operatorname{Re}\left\langle \mathbf{E}\right\rangle _{\Psi}\\
x_{4} &  =\operatorname{Im}\left\langle \mathbf{E}\right\rangle _{\Psi
}.\nonumber
\end{align}

Note that $\left\langle \mathbf{E}\right\rangle _{\Psi}$ vanishes for non
entangled states, and takes its maximal norm (equals to 1) for maximally
entangled states, hence the name ''entanglor''. \ Such an operator, which is
nothing but the time reversal operator for two spins
${\frac12}$%
, has been already largely used in quantifying entanglement\cite{wootters},
with a quantity $c$, called the ''concurrence'', which here equals $c=2$
$\left|  C_{2}\right|  $.

\subsection{Discussion}

Let us now see what has been gained in fibrating the \ $\mathcal{E}$ two-qubit
Hilbert space. It is clear that, at some point, we will need to focus on the
''projective'' Hilbert space, taking into account the global phase freedom. Is
is nevertheless interesting to first stay at the Hilbert space level, and
foliate the base space along constant values of simple operators expextation
values. The simplest \ thing to do is to fix \ $\left\langle \sigma_{z}\otimes
Id\right\rangle _{\Psi}$, e.g. $x_{0}$. Let us start from the $S^{4}$ base
north pole, $x_{0}=1$, which is the image of the states%
\begin{equation}
\Psi_{Q=\infty}=\left|  0\right\rangle _{1}\otimes\left(  \alpha\left|
0\right\rangle _{2}+\beta\left|  0\right\rangle _{2}\right)
\end{equation}

Decreasing the $x_{0}$ values, we get a sequence of ''horizontal'' $S^{3}$
spheres of radius $\sin2\Omega$, denoted $S_{\Omega}^{3}$. This is nothing but
the generalization in higher dimension of the standard horizontal ''parallel''
circles which foliates $S^{2}$. As is well known, $S^{3}$ can also be foliated
by nested tori, with two opposite nested great circles sitting on the
orthogonal two planes $\left(  x_{1},x_{2}\right)  $ and $\left(  x_{3}%
,x_{4}\right)  $. As already noticed, states such $\ x_{3}=x_{4}=0$ are
non-entangled states. Maximally entangled states ($M.E.S.$) correspond to
$x_{0}=x_{1}=x_{2}=0$. As far as the geometric description is concerned, it is
also usefull to define $\Omega.M.E.S.$, which are states with fixed values of
$\left\langle \sigma_{z}\otimes Id\right\rangle _{\Psi}$ (e.g., constant
$x_{0}$) and, with this constraint, maximal entanglement (e.g., maximal
concurrence). Usual $\ M.E.S$ are nothing but $\frac{\pi}{4}.M.E.S$, and we
show below that all $\Omega.M.E.S$ manifolds share the same topology, except
for the extremal values $\Omega=0$ and $\Omega=\pi/2$, which are separable states.

Let us now inverse the Hopf map, and get the general expression for a state (a
$S^{7}$ point ) which is sent to $Q$ by the $h_{1}$ map. Such a state, noted
$\Psi_{Q}$ reads (given as a pair of quaternions)
\begin{equation}
\Psi_{Q}=(\cos\Omega\exp\left(  -\theta\mathbf{t}/2\right)  \,q\,,\sin
\Omega\exp\left(  \theta\mathbf{t}/2\right)  \,q),
\end{equation}

where $\cos\theta=x_{1}/\sin2\Omega=S(Q^{\prime})$, $q$ is a unit quaternion
which spans the $S^{3}$ fibre and $\mathbf{t}$ is the following unit pure
imaginary quaternion :%
\begin{equation}
\mathbf{t=}\left(  \mathbf{V}_{\mathbf{i}}(Q^{\prime})\mathbf{i}%
+\mathbf{V}_{\mathbf{j}}(Q^{\prime})\mathbf{j}+\mathbf{V}_{\mathbf{k}%
}(Q^{\prime})\mathbf{k}\right)  /\sin\theta\text{. }%
\end{equation}

\subsubsection{Separable states}

In the non entangled case, $Q$ is now a complex number and the above
expression simplifies to%
\begin{equation}
\Psi_{Q}=(\cos\Omega\exp\left(  -\theta\mathbf{i}/2\right)  \,q,\sin\Omega
\exp\left(  \theta\mathbf{i}/2\right)  \,q).
\end{equation}

The projective Hilbert space for two non-entangled qubits $A$ and $B$ is
expected to be the product of two 2-dimensional spheres $S_{A}^{2}\times
S_{B}^{2}$, each sphere being the Bloch sphere associated with the given
qubit. This property is clearly displayed here. The unit $S^{4}$ base space
reduces in a unit $S^{2}$ sphere (since $x_{3}=x_{4}=0)$ which is nothing but
the Bloch sphere for the first qubit, as seen on the coordinate expressions,
$\left\langle \sigma_{z,x,y}\otimes Id\right\rangle _{\Psi}$, for $x_{0}%
,x_{1}$ and $x_{2}$ respectively. This $S^{2}$ sphere is the aggregate of
parallel circles, of radius $\sin2\Omega$ in the plane $\left(  x_{1}%
,x_{2}\right)  $, each one on a parallel $S_{\Omega}^{3}$ sphere, and with the
extremal poles corresponding to $\Omega=0$ and $\Omega=\pi/2$. For the second
qubit Bloch sphere, one needs to define a coordinate system on the fibres. We
choose the two orthogonal states $\left|  0\right\rangle _{Q}$ and $\left|
1\right\rangle _{Q}$ respectively corresponding to $q=1$ and $q=j$ in the
above expression for $\Psi_{Q}$. They read, in the initial $\mathcal{E}$
product state basis%
\begin{align}
\left|  0\right\rangle _{Q} &  =\left(  \cos\Omega\exp\left(  -i\theta
/2\right)  \,\left|  0\right\rangle _{1}+\sin\Omega\exp\left(  i\theta
/2\right)  \,\left|  1\right\rangle _{1}\right)  \otimes\left|  0\right\rangle
_{2}\\
\left|  1\right\rangle _{Q} &  =\left(  \cos\Omega\exp\left(  -i\theta
/2\right)  \,\left|  0\right\rangle _{1}+\sin\Omega\exp\left(  i\theta
/2\right)  \,\left|  1\right\rangle _{1}\right)  \otimes\left|  1\right\rangle
_{2}\nonumber
\end{align}

A generic state $\Psi_{Q}$ in the $S^{3\text{ }}$fibre reads $\left|  \Psi
_{Q}\right\rangle =a\left|  0\right\rangle _{Q}+b\left|  1\right\rangle _{Q},$
with $a,b\in\mathbb{C}\mathbf{,\quad}\left|  a\right|  ^{2}+\left|  b\right|
^{2}=1.$ We can now iterate the fibration process on the fibre itself and get
the (Hopf fibration base)-(Bloch sphere) coordinates for this two-level
system.\ It is now a simple exercise to verify that the obtained 3 real
coordinates are precisely $\left\langle Id\otimes\sigma_{x,y,z}\right\rangle
_{\Psi}$, that is to say the second qubit Bloch sphere.

In summary, for non entangled qubits, the $S^{7}$ Hopf fibration, with base
$S^{4}$ and fibre $S^{3}$, simplifies to the simple product of a $S^{2}$
sub-sphere of the base (the first qubit Bloch sphere) by a second $S^{2}$ (the
second qubit Bloch sphere) obtained as the base of a $S^{3}$ Hopf fibration
applied to the fibre itself. Let us stress that this last iterated fibration
is necessary to take into account the global phase of the two qubit system.

The fact that these two $S^{2}$ spheres play a symmetrical role (although one
is related to the base and the other to the fibre) can be understood in the
following way. We grouped together $\alpha$ and $\beta$ on one hand, and
$\gamma$ and $\delta$ on the other hand, to form the quaternions $q_{1}$ and
$q_{2}$, and then define the Hopf map $h_{1}$ as the ratio of these two
quaternions( plus a complex conjugation). Had we grouped $\alpha$ and $\gamma
$, and $\beta$ and $\delta$, to form two new quaternions, and use the same
definition for the Hopf map, we would also get a $S^{7}$ Hopf fibration, but
differently oriented. We let as an exercise to compute the base and fibre
coordinates in that case. The net effect is to interchange the role of the two
qubits: the second qubit Bloch sphere is now part of the $S^{4}$ base, while
the first qubit Bloch sphere is obtained from the $S^{3}$ fibre.

\subsubsection{Entangled states}

As expected, the entangled case is more complicated. From an analytical point
of view, this is related to the fact that the pure imaginary quaternion
$\mathbf{t}$ doesnot reduce to the standard $\mathbf{i}$ imaginary unit. Let
us focus first on the maximally entangled states. They correspond to the
complex number $C_{2}$ having maximal norm $1/2$. This in turn implies that
the Hopf map base reduces to a unit circle in the plane $\left(  x_{3}%
,x_{4}\right)  $, parametrized by the unit complex number $2C_{2}$. The
projective Hilbert space for these M.E.S. is known to be $S^{3}/Z_{2},$ a
$S^{3}$ \ sphere with opposite points identified \cite{2qbits_geometry} (this
is linked to the fact that all M.E.S. can be related by a local operation on
one sub-system, since $S^{3}/Z_{2}=SO(3)$). In order to recover this result in
the present framework, one can follow the trajectory of a representative point
on the base and on the fibre while the state is multiplied by an overall phase
$\exp\left(  i\varphi\right)  $. The expression for $C_{2}\left(
=\alpha\delta-\beta\gamma\right)  $ shows that the point on the base turns by
twive the angle $\varphi$. Only when $\varphi=\pi$ does the corresponding
state belongs to the same fiber (e.g. maps onto the same value on the base).
The fact that the fibre is a $S^{3}$ sphere, and this two-to-one
correspondance between the fibre and the base under a global phase change,
explains the $S^{3}/Z_{2}$ topology for the $M.E.S.$ projective Hilbert space.

In that case, $\Psi_{Q}$ reads%
\begin{equation}
\frac{1}{\sqrt{2}}(\exp\left(  -\pi C_{2}\mathbf{j}/2\right)  \,q\,,\exp
\left(  \pi C_{2}\mathbf{j}/2\right)  \,\,q).
\end{equation}

The standard four Bell states are recovered with $q=\left(  1\pm
\mathbf{j}\right)  /\sqrt{2}$ and $C_{2}=\pm%
\frac12
$

Now, it is clear that the previous discussion can be repeated for any
$\Omega.M.E.S.$, which have therefore the same $S^{3}/Z_{2}$ topology, and
reads%
\begin{equation}
(\cos\Omega\exp\left(  -\frac{\pi}{4}\frac{C_{2}}{\left|  C_{2}\right|
}\mathbf{j}\right)  \,q\,,\sin\Omega\exp\left(  \frac{\pi}{4}\frac{C_{2}%
}{\left|  C_{2}\right|  }\mathbf{j}\right)  \,\,q).
\end{equation}

Finally, we consider generic two-qubit states, which are neither separable nor
maximally entangled. For each value of $\Omega$ (e.g., of $\left\langle
\sigma_{z}\otimes Id\right\rangle _{\Psi}$), these states are mapped (through
the $h_{1}$ map) onto a torus of the $S_{\Omega}^{3}$ torus foliation. Such a
torus reduces to a circle of radius $2\left|  C_{1}\right|  $ for the
projective version. To each point of these circles is again associated a
$S^{3}/Z_{2}$ manifold in the fibre part.

\subsubsection{A tentative generalization of the Bloch sphere representation}

We are now led to consider a tentative (and partly well known, see below)
generalization of the Bloch sphere for the two-qubit projective Hilbert space.
Clearly, the present Hopf fibration description suggests a splitting of the
representation space in a product of base and fibres sub-spaces. Of the base
space $S^{4}$, we propose to only keep the first three coordinates
\begin{equation}
\left(  x_{0},x_{1},x_{2}\right)  =\left(  \left\langle \sigma_{z}\otimes
Id\right\rangle _{\Psi},\left\langle \sigma_{x}\otimes Id\right\rangle _{\Psi
},\left\langle \sigma_{y}\otimes Id\right\rangle _{\Psi}\right)
\end{equation}

All states map into a standard ball $B^{3}$ of radius 1, where the set of
separable states forms the $S^{2}$ boundary (the usual first qubit Bloch
sphere), and the centre corresponds to maximally entangled states. Concentric
spherical shells around the centre correspond to states of equal concurrence
$c$ (maximal at the centre, zero on the surface), the radius of the spherical
shell being equal to $\sqrt{1-c^{2}}$. The idea of slicing the 2-qubit Hilbert
space into manifolds of equal concurrence is not new \cite{2qbits_geometry}.
What is nice here is that, under the Hopf map (and a projection onto the 3d
subspace of the base spanned by the first three coordinates), these manifolds
transforms into concentric $S^{2}$ shells which fill the unit ball. We must
now describe which subset of the $S^{3}$ fibres has to attached to each point
in this base-related unit ball. Looking to the previous discussions, it is
clear that it is a $S^{3}/Z_{2}$ space for each interior point, and a $S^{2}$
space (the second qubit Bloch sphere) for the boundary of the unit ball.

The Bloch ball single qubit mixed state representation was recalled above. The
centre of the Bloch ball corresponds to maximally mixed states. The reader
might be surprised to find here (in the two-qubit case) a second unit radius
ball, with maximally entangled states at the centre.There is no mystery here,
and it corresponds to a known relation between partially traced two-qubit pure
states and one-qubit mixed state. Indeed the partially traced density matrix
$\rho_{1}$ is in fact deeply related to the $S^{7}$ Hopf map:%

\begin{equation}
\rho_{1}=\left(
\begin{array}
[c]{cc}%
\left|  q_{1}\right|  ^{2} & \overline{C_{1}}\\
C_{1} & \left|  q_{2}\right|  ^{2}%
\end{array}
\right)  =\frac{1}{2}\left(
\begin{array}
[c]{cc}%
1+x_{0} & x_{1}-\mathbf{i\,}x_{2}\\
x_{1}+\mathbf{i\,}x_{2} & 1-x_{0}%
\end{array}
\right)  \text{ \ }%
\end{equation}

with unit trace and $\det\rho_{1}=\left|  C_{2}\right|  ^{2}.$ The partial
$\rho_{1}$ represents a pure state density matrix whenever $C_{2}$ vanishes
(the separable case), and allows for a unit Bloch sphere (that associated to
the first qubit). It is assimilated to a mixed state density matrix as soon as
$\left|  C_{2}\right|  >0$ (and an entangled state for the two qubit state).
The other partially traced density matrix $\rho_{2}$ is related to the other
$S^{7}$ Hopf fibration which was discussed above. With all this in mind, one
also expect a simple generic relation between Hopf map and the Schmidt
decomposition. Indeed the two common eigenvalues of $\rho_{1}$and $\rho_{2}$
read%
\begin{equation}
\lambda_{\pm}=\frac{1+\sqrt{1-4\left|  C_{2}\right|  ^{2}}}{2}=\frac
{1+\sqrt{1-c^{2}}}{2}%
\end{equation}

from which we get the Schmidt weights $\cos\frac{\varepsilon}{2}=\sqrt
{\lambda_{+}}$ and sin$\frac{\varepsilon}{2}=\sqrt{\lambda_{-}}$. The Schmidt
states are easilly obtained from the diagonalization of $\rho_{1}$ and
$\rho_{2}$ . Their decomposition in the original product state basis can
therefore also be given in term of the Hopf map parameters. As an an example,
the two (unnormalized) states for the first qubit simply read%
\begin{equation}
\left|  \varphi_{\pm}\right\rangle =\overline{C_{1}}\left|  0\right\rangle
_{1}+\left(  \lambda_{\pm}-\left|  q_{1}\right|  ^{2}\right)  \left|
1\right\rangle _{1}%
\end{equation}

\section{Conclusion}

Our main goal in this paper was to provide a geometrical representation of the
two-qubit Hilbert space pure states. We hope to have convinced the reader that
the $S^{7}$ Hopf fibration plays a natural role in that case: it provides a
adapted ''skeletton'' for preparing a meaningfull slicing of the full Hilbert
space. It is rather fascinating that the two main Hopf fibrations (of $S^{3}$
and $S^{7}$) are so nicely related to the two important ingredients of the
quantum objects, which are the global phase freedom and entanglement.

In going further, one may ask two natural questions. The first concerns the
geometry of two-qubit mixed states, and the second the generalization to more
than two qubits. Note that these two questions are not necessarily completly
independant, as we just saw in the previous paragraph. Up to now, we have
mainly focused on the second question, and addressed the 3-qubit
case\cite{mosseri}. Indeed, the Hilbert space is now $S^{15}$, which also
allow for a Hopf fibration, with a $S^{8}$ base and $S^{7}$ fibres. From an
analytical point of view, the Hopf map takes the same form, but quaternions
need to be replaced by ''octonions''. This is not a harmless change, since
octonion algebra is not only non-commutative, but also non-associative
(forbiding a matrix representation for instance). But this analysis can be
done, and provides interesting insights in the Hilbert space geometry. Note
that in the three Hopf fibrations sequence ($S^{15},S^{7},S^{3})$, the fibre
in the larger dimensional space is the full space in next case. This offers
the possibility of further nesting the fibrations, a possibility that we
already used in the present analysis of two-qubit separable case.

\begin{center}
\textbf{Appendix: Quaternions}
\end{center}

Quaternions are usually presented with the imaginary units $\mathbf{i}%
,\mathbf{j}$ et $\mathbf{k}$ in the form~:%
\[
q=x_{0}+x_{1}\mathbf{i}+x_{2}\mathbf{j}+x_{3}\mathbf{k},\qquad x_{0}%
,x_{1},x_{2},x_{3}\in\mathbb{R}\quad with\mathbf{i}^{2}=\mathbf{j}%
^{2}=\mathbf{k}^{2}=\mathbf{ijk}=-1,
\]

the latter ''Hamilton'' relations defining the quaternion multiplication rules
which are non-commutative. They can also be defined equivalently, using the
complex numbers $c_{1}=x_{0}+x_{1}\mathbf{i}$ and $c_{2}=x_{2}+x_{3}%
\mathbf{i}$, in the form $q=c_{1}+c_{2}\mathbf{j}$.

The conjugate of a quaternion $q$ is $\overline{q}=x_{0}-x_{1}\mathbf{i}%
-x_{2}\mathbf{j}-x_{3}\mathbf{k=}\overline{c_{1}}-c_{2}\mathbf{j}$ and its
squared norm reads $N_{q}^{2}=q\overline{q}$.

Another way in which $q$ can be written is as a scalar part $S(q)$ and a
vectorial part $\mathbf{V}(q)$:%
\[
q=S(q)+\mathbf{V}(q),\;S(q)=x_{0},\;\mathbf{V}(q)=x_{1}\mathbf{i}%
+x_{2}\mathbf{j}+x_{3}\mathbf{k,}%
\]

with the relations%
\[
S(q)=\frac{1}{2}(q+\overline{q}),\;\mathbf{V}(q)=\frac{1}{2}(q-\overline{q}).
\]

A quaternion is said to be real if $\mathbf{V}(q)=0$, and pure imaginary if
$S(q)=0$. We shall also write$\;\mathbf{V}_{\mathbf{i},\mathbf{j},\mathbf{k}%
}(q)$ for the component of $\mathbf{V}(q)$ along $\mathbf{i},\mathbf{j}%
,\mathbf{k.}$ Finally, and as for complex numbers, a quaternion can be noted
in an exponential form as%
\[
q=\left|  q\right|  \exp\varphi\mathbf{t=}\left|  q\right|  \left(
\cos\varphi+\sin\varphi\quad\mathbf{t}\right)  \text{, }%
\]

where $\mathbf{t}$ is a unit pure imaginary quaternion. When $\mathbf{t}%
=\mathbf{i}$, usual complex numbers are recovered. Note that quaternion
multiplication is non-commutative so that%
\[
\exp\varphi\mathbf{t}\exp\lambda\mathbf{u=\exp(}\varphi\mathbf{t+}%
\lambda\mathbf{u)}%
\]

only if $\mathbf{t=u}$.

\bigskip

\textbf{Acknowledgement}

It is a pleasure for one of the authors (R.M.) to acknowledge enlightning
discussions about quantum information with O. Cohen and B. Griffiths.

The content of this paper was presented at the national congress of the French
Physical Society, Strasbourg, 9-13 july 2001 (unpublished).

While completing the present written version, we met a very interesting
preprint by I. Bengtsson, J. Br\"{a}nnlund and K. Zyczkowski
(quant-phys/0108064) on the geometry of qubits projective Hilbert space, with
$S^{3}$ Hopf fibrations occasionaly used to describe submanifolds. Beside a
different perspective in slicing the Hilbert space, a main difference with our
work is the present use of the $S^{7}$ Hopf fibration applied to the whole space.


\begin{thebibliography}{99}
\bibitem{q_info_gen}D. Bouwmeester, A. Eckert, A. Zeilinger, \textit{The
Physics of Quantum Information, }Springer-Verlag \ 2000

\bibitem {2qbits_geometry}M. Kus and K. Zyczkowski, Phys. Rev A. \textbf{63}
(2001) page 032307, and references herein.

\bibitem {urtbanke}H. Urbanke, American Journal of Physics, \textbf{59} (1991)
page 53

\bibitem {sadoc_mosseri_livre}J.F. Sadoc and R. Mosseri, \textit{Geometric
Frustration}, Cambridge University Press 1999)

\bibitem {peres}A. Peres, \textit{Quantum Theory: Concepts and methods},
Kluwer, Dordrecht 1993

\bibitem {wootters}W.K. Wootters, Phys. Rev. Lett. \textbf{80} (1998) 2245

\bibitem {mosseri}R. Mosseri, in preparation
\end{thebibliography}
\end{document}